\input phyzzx
\input tables
\voffset = -0.4in
\footline={\ifnum\pageno=1 \nulline \else\newfootline \fi}
\def\nulline{{\hfill}}
\def\newfootline{\advance\pageno by -1\hss\tenrm\folio\hss}
\rightline {Oct 1995}
\rightline {QMW--PH--96--1 }
\rightline {SUSX--TH--96--004 }
\title {GAUGE COUPLING CONSTANT UNIFICATION WITH PLANCK SCALE VALUES OF MODULI}
\author{D. Bailin$^{a*}$, \ A. Love$^{b}$,  \ W.A.
Sabra$^{b**}$\ and \ S. Thomas$^{c***}$}
\address {$^{a}$School of Mathematical and Physical
Sciences,\break
University of Sussex, \break Brighton U.K.}
\address {$^{b}$Department of Physics,\break
Royal Holloway and Bedford New College,\break
University of London,\break
Egham, Surrey, U.K.}
\address {$^{c}$
Department of Physics,\break
Queen Mary and Westfield College,\break
University of London,\break
Mile End Road, London,  U.K.}
\footnote*{e-mail address: D.BAILIN@SUSSEX.AC.UK.}
\footnote{**}{e-mail address: UHAP012@VAX.RHBNC.AC.UK.}
\footnote{***}{e-mail address: S.THOMAS@QMW.AC.UK.}
\abstract {Convergence of the standard model gauge coupling constants
to a common value at around $2\times 10^{16}$ GeV is studied in the context
of orbifold theories where the modular symmetry groups for $T$ and $U$
moduli are broken to subgroups of $PSL(2, Z)$. The values of the moduli
required  for this unification of coupling constants are studied for this case
and also for
the case where string unification is accompanied by unification to a gauge
group larger then
$SU(3)\times SU(2)\times U(1).$}
\endpage
\REF\one {J. Ellis, S. Kelley and D. V. Nanopoulus, {\it Phys. Lett.} {\bf
B260} (1991) 131.}
\REF\two{ U. Amaldi, W. de Boer and H. Furstenau,  {\it Phys. Lett.} {\bf
B260} (1991) 447.}
\REF\three{P. Ginsparg, {\it Phys. Lett.} {\bf B197} (1987) 139.}
\REF\four {L. E. Ibanez , {\it Phys. Lett.} {\bf B126} (1983) 196;
J. E. Bjorkman and D. R. T. Jones, {\it Nucl. Phys.} {\bf B259} (1985) 533.}
\REF\five { I. Antoniadis, J. Ellis, S. Kelley  and D. V. Nanopoulus, {\it
Phys.
Lett.} {\bf B271} (1991) 31.}
\REF\six{ D. Bailin and A. Love, {\it Phys. Lett.} {\bf  B280} (1992) 26;
D. Bailin and A. Love,  {\it Mod. Phys. Lett.} {\bf A7}  (1992) 1485.}
\REF\seven{ S. Kelley, J. L. Lopez and and D. V. Nanopoulus, {\it Phys. Lett.}
{\bf  B278} (1992) 140.}
\REF\dienesa{K. R. Dienes and A. E. Faraggi, {\it Phys. Rev. Lett.} {\bf 75}
(1995) 2646; K. R. Dienes and A. E. Faraggi, {\it Nucl. Phys.} {\bf B457}
(1995) 409.}
\REF\eight{V. S. Kaplunovsky, {\it Nucl. Phys}. {\bf B307} (1988) 145.}
\REF\nine {L. J. Dixon, V. S. Kaplunovsky and J. Louis,  {\it Nucl. Phys}.
{\bf B355} (1991) 649.}
\REF\ten{I. Antoniadis, K. S. Narain and T. R. Taylor,
 {\it Phys. Lett}.
{\bf B267} (1991) 37. }
\REF\eleven{J. P. Derenddinger, S. Ferrara, C. Kounas and F. Zwirner, {\it
Nucl. Phys.} {\bf B372} (1992) 145.}
\REF\twelve{D. Bailin and A. Love, {\it Phys. Lett.} {\bf B292} (1992) 315.}
\REF\dienesb{K. R. Dienes,  A. E. Faraggi and J. March-Russell, hep-th/9510223,
to appear in {\it Nucl. Phys.} {\bf B}.}
\REF\thirteen{D. Lewellen, {\it Nucl. Phys}.
{\bf B337} (1990) 61.}
\REF\fourteen{ A. Font, L. E. Ibanez,
and F. Quevedo, {\it Nucl. Phys}. {\bf B345} (1990) 389.}
\REF\fifteen {L. E. Ibanez, D. Lust and G. G. Ross,  {\it Phys. Lett}.
{\bf B272} (1991) 251.}
\REF\sixteen {L. E. Ibanez and  D. Lust , {\it Nucl. Phys}. {\bf B382}
(1992) 305.}
 \REF\seventeen{ P. Mayr and S. Stieberger, {\it Nucl. Phys}. {\bf B407}
(1993)
725.}
\REF\eighteen{D. Bailin, A. Love, W. A. Sabra and S. Thomas, {\it Mod. Phys.
Lett}. {\bf A 9} (1994) 67; {\it Phys. Lett.} {\bf  B320} (1994) 21.}
\REF\nineteen{D. Bailin, A. Love, W. A. Sabra and S. Thomas, {\it Mod. Phys.
Lett}. {\bf A 9} (1994) 1229; D. Bailin, A. Love, W. A. Sabra and S. Thomas,
{\it Nucl. Phys}. {\bf B427}
(1994) 181.}
\REF\twenty{M. Spalinski,  {\it Phys.  Lett}. {\bf B275} (1992) 47;
J. Erler, D. Jungnickel and H. P. Nilles, {\it Phys. Lett.} {\bf B276} (1992)
303;
J. Erler and M. Spalinski,
{\it Int J  Mod. Phys.}. {\bf A 9} (1994) 4407.}
\REF\twentyone{ H. P. Nilles and S. Stieberger, hep-th/951009.}
\REF\twentytwo{ D. Bailin, A. Love, W. A. Sabra and S. Thomas,  {\it Mod. Phys.
Lett.} {\bf A9} (1994) 2543.}
\REF\twentythree{D. Bailin and A. Love,  {\it Phys. Lett}.
{\bf B288} (1992) 263.}
\REF\twentyfour{B. R. Greene, K. H. Kirklin, P. J. Miron and G. G. Ross,
{\it Phys. Lett}.
{\bf B180} (1986) 69.}
\REF\twentyfive{D. Bailin, A. Love, W. A. Sabra and S. Thomas,  {\it Mod. Phys.
Lett.} {\bf A10} (1995) 337.}
\REF\twentysix{I. Antoniadis and G. K. Leontaris,  {\it Phys. Lett}.
{\bf B216} (1989) 333.}

When the standard model gauge coupling constants are extrapolated [\one, \two]
to high energies using the renormalization group equations of the minimal
supersymmetric model with just two Higgs doublets the three gauge couplings
$g_3$, $g_2$ and $g_1$ of $SU(3)\times SU(2)\times U(1)$ attain a common value
at about $2\times 10^{16}$ GeV.
There is a problem in obtaining consistency with heterotic string theory
because tree level gauge coupling constants in the string theory have a common
value [\three] at a string unification scale $M_{string}$ around $0.37\times
10^{18}$ GeV. Amongst the possible ways of arranging consistency are grand
unification of the gauge group to $SU(5)$ or $SO(10)$ at $2\times 10^{16}$ GeV
with the coupling constants then running with a common value to $M_{string},$
modification of the running of the renormalization group equations by the
inclusion of extra states [\four-\dienesa] with mass intermediate between the
electroweak scale and the string unification scale, and, in the context of
orbifold compactification, inclusion of moduli dependent string loop threshold
corrections [\eight-\eleven] in the renormalization group equations for the
standard model or for models with $SU(5)\times U(1)$, $SO(4)\times SO(6)$ or
$[SU(3)]^3$ unification [ \twelve].  Unification of gauge coupling
constants has also been studied in the context of free fermion models
[\dienesa ]. Furthermore, an alternative approach in which one considers
non-standard values of the Kac-Moody levels within the minimal supersymmetric
standard model has been studied in [\dienesb].

The first of these approaches requires the gauge group of the heterotic string
theory to be at least at level two to permit Higgs scalars in the adjoint
representation [\thirteen]. It has proved difficult to construct realistic
models of this type despite considerable efforts [\fourteen]. The second
approach requires us to believe that the observed unification of gauge coupling
constants at $2\times 10^{16}$ GeV using supersymmetric standard model
renormalization group equations is a coincidence, and without unification to a
gauge group larger than $SU(3)\times SU(2)\times U(1),$ the third approach
appears to require large values of the orbifold moduli to give a sufficiently
large threshold corrections [\fifteen,\sixteen]. However, it has been assumed
in the latter calculations that the threshold corrections are those with
$PSL(2, Z)$ modular symmetry in the $T$ and $U$ moduli.
These modular symmetry groups can be broken [\seventeen, \eighteen] to
subgroups of $PSL(2, Z)$ when the orbifold lattice is such that some twisted
sectors have fixed planes for which the six-torus ${\bf T}_6$ cannot be
decomposed into a direct sum ${\bf T}_2 \oplus{\bf T}_4$ with the fixed plane
lying in  ${\bf T}_2$.  We shall refer to this as the non ${\bf T}_2\oplus{\bf
T}_4$ case. The modified form of the threshold corrections is known
[\seventeen, \eighteen].
Modular symmetries of threshold corrections may also be broken by discrete
Wilson line backgrounds [\sixteen, \nineteen, \twenty] though in this case it
has not been possible to date to calculate the form of the threshold
corrections.

We shall investigate the effect of threshold corrections with broken modular
symmetries on the values of the $T$ and $U$ moduli for which unification of
gauge group couplings constants occurs at $2\times 10^{16}.$
( In a recent paper [\twentyone] it has been shown that the inclusion of Wilson
line moduli along with $T$ and $U$ moduli can result in substantially smaller
values of moduli being needed.)
We shall also study the values of the moduli required to achieve this
unification of gauge coupling constants when the gauge group above the
unification scale is larger than $SU(3)\times SU(2)\times U(1)$.

In general, the renormalization group equations, including string loop
threshold corrections, for a semi-simple gauge group with factors $G_a$, all at
level 1, may be written in the form
$$16\pi^2g_a^{-2}(\mu)=16\pi^2g_{string}^{-2}+b_a log
\Big({M^2_{string}\over\mu^2}\Big)+\Delta_a,\eqn\on$$
where $g_{string}$ is the common value of the gauge coupling constants at the
string tree level unification scale $M_{string}$ with approximate values
$$M_{string}\approx 0.527 g_{string}\times 10^{18} \hbox{GeV}\eqn\tw$$
and $$g_{string}\approx 0.7\eqn\thre$$

In the non ${\bf T}_2\oplus {\bf T}_4$ cases, with modular symmetries that are
subgroups of $PSL(2, Z)$, the threshold corrections take the form [\seventeen,
\eighteen]
$$\eqalign{\Delta_a=&-\sum_i({{b'}_a}^i-\delta_{GS}^i)\Big(ln({ T}_i+\bar {
T}_i)+\sum_m{C_{im}\over2}ln \vert\eta({{ T}_i\over l_{im}} ) \vert^4\Big)\cr
-&\sum_i({{d'}_a}^i-{\tilde\delta}_{GS}^i)\Big(ln(U_i+\bar U_i)+
\sum_m{{\tilde C}_{im}\over2}ln \vert\eta({U_i\over {\tilde l}_{im}} )
\vert^4\Big)}\eqn\fou$$
where the sum over $i$ is restricted to $N=2$ complex planes, which are
unrotated in at least one twisted sector, and for the $U$ moduli is further
restricted to complex planes for which the point group acts as $Z_2$.
The range over which $m$ runs depends on the value of $i$, but
$$\sum_m {\tilde C}_{im}=\sum_m {C}_{im}=2,\qquad \forall i, \eqn\fiv$$
The coefficients ${C}_{im}$, ${l}_{im}$, ${\tilde C}_{im}$ and
${\tilde l}_{im}$ are given in [\twentytwo] for the various non ${\bf
T}_2\oplus{\bf T}_4$ Coxeter ${\bf Z}_N$ orbifolds. In the case of the ${\bf
Z}_6-II-b$ orbifold, the modulus $U_3$ is understood to be replaced by $U_3+2i$
in the argument of the Dedekind eta function.
The quantities ${\delta}_{GS}^i$ and ${\tilde\delta}_{GS}^i$ are the
Green-Schwarz coefficients, and the coefficients
${{b'}_a}^i$ and ${{d'}_a}^i$, which are determined by the contribution of the
massless states to the modular anomaly [\eleven] in a way that does not depend
on the underlying lattice of the orbifold are given by
$$ {{b'}_a}^i=-C(G_a)+\sum_{R_a}T(R_a)(1+2n_{R_a}^i)\eqn\si$$
and
$$ {{d'}_a}^i=-C(G_a)+\sum_{R_a}T(R_a)(1+2l_{R_a}^i)\eqn\seve$$
where $C(G_a)$ and $T(R_a)$ are Casimirs for the gauge group factor $G_a$
and its representations $R_a$, and $n_{R_a}^i$ and $l_{R_a}^i$ are the
modular weights under ${T}_i$ and $U_i$ modular transformations, respectively,
for massless states in the representation $R_a$ of $G_a$.
All possible values of $n_{R_a}^i$ and $l_{R_a}^i$ have been determined
[\sixteen, \twentythree] for massless states in arbitrary twisted sectors of
abelian
Coxeter orbifolds.

If $g_a$ and $g_b$ are the gauge coupling constants for 2 factors of the
$SU(3)\times SU(2)\times U(1)$ standard model gauge group, and if the
unification scale at which all 3 gauge coupling constants converge to a
common value $M_X$ with
$$M_X=2\times 10^{16}\eqn\eigh$$
then, from \on\ and \fou,
$${M^2_{string}\over M_X^2}=\prod_i\alpha_i^{{({{b'}_a}^i-{{b'}_b}^i)}\over
{({{b}_a}-{{b}_b})}}  {\tilde\alpha}_i^{{({{d'}_a}^i-{{d'}_b}^i)}\over
{({{b}_a}-{{b}_b})}}\eqn\nin$$
where $$\alpha_i=({T}_i+\bar {T}_i)\prod_m\vert\eta({{ T}_i\over
l_{im}})\vert^{2C_{im}}\eqn\te$$
and
$${\tilde\alpha}_i=(U_i+\bar U_i)\prod_m\vert\eta({U_i\over {\tilde
l}_{im}})\vert^{2{\tilde C}_{im}}\eqn\eleve$$

In the case of $\alpha_i$, the product in \nin\ sums over all $N=2$ complex
planes, and in the case of ${\tilde\alpha}_i$ over all $N=2$ complex planes for
which the point group acts as $Z_2$.

For the supersymmetric standard model with $SU(3)\times SU(2)\times U(1)$
gauge group, 3 generations of quarks and leptons and higgses $h$ and
$\bar h$, the renormalization group coefficients $b_a$ are
$$b_3=-3,\  b_2=1, \  b_1={33\over5}.\eqn\twelv$$
In terms of the modular weights for the massless matter fields, the
coefficients  ${{b'}_a}^i$ are given by
$${{b'}_3}^i=3+\sum_{g=1}^3(2n_{Q(g)}^i+n_{u(g)}^i+n_{d(g)}^i)\eqn\thirtee$$
$${{b'}_2}^i=5+n_h^i+n_{\bar
h}^i+\sum_{g=1}^3(3n_{Q(g)}^i+n_{L(g)}^i)\eqn\fourtee$$
and
$${{b'}_1}^i={33\over5}+{3\over5}(n_h^i+n_{\bar
h}^i)+{1\over5}\sum_{g=1}^3(n_{Q(g)}^i+8n_{u(g)}^i+2n_{d(g)}^i+3n_{L(g)}^i+6n_{e(g)}^i)\eqn\fiftee$$
where $g$ labels the generations, and $L(g)$ and $Q(g)$ are lepton and
quark $SU_L(2)$ doublets. Exactly similar expressions apply for ${{d'}_a}^i$
with $n^i$ replaced by $l^i$.

For a given twisted sector of a given orbifold the possible modular
weights of matter states can be calculated from the twists on the string
degrees of freedom and the left mover oscillators involved in the
construction of the states [\sixteen, \twentythree]. In general, for a
massless left mover the oscillator number $\tilde N$ is given by
$$\tilde N= a_L-h_{KM}\eqn\sixtee$$
where $a_L$ is the normal ordering constant for the particular orbifold
twisted sector and $h_{KM}$ is the contribution to the conformal weight
of the state from the $E_8\times E_8$ algebra. For level 1 gauge group
factors $G_a$, the contribution is given by [\sixteen]
$$h_{KM}=\sum_a {dim G_a\over dim R_a}{T(R_a)\over C(G_a)+1}\eqn\seventee$$

For the $SU(3)\times SU(2)\times U(1)$ case, the relevant conformal
weights are
$$ h_{KM}\ge {3\over5},\quad \hbox{for}\ Q, u, e,\eqn\eightee$$
and
$$ h_{KM}\ge {2\over5},\quad \hbox{for}\  L,  d, h, \bar h,\eqn\ninetee$$
where the inequality allows for any additional contributions to
$h_{KM}$ from extra $U(1)$ factors in the gauge group assumed to be
spontaneously broken along flat directions at a high energy scale.

Because the complex planes for which both $T$ and $U$ moduli occur are planes
where the point group acts as $Z_2$, the modular weights associated with
the $T$ and $U$ modulus for these planes are the same state by state.
As a consequence, for such complex planes we have ${{b'}_a}^i$ and
${{d'}_a}^i$ equal. Thus \nin\ simplifies to
$${M^2_{string}\over M_X^2}=\prod_j\alpha_j^{{({{b'}_a}^j-{{b'}_b}^j)}\over
{({{b}_a}-{{b}_b})}} \prod_k({{\alpha}_k {\tilde\alpha}_k})^
{{({{b'}_a}^k-{{b'}_b}^k)}\over
{({{b}_a}-{{b}_b})}}\eqn\twent$$
where the product over $k$ is for the $N=2$ $Z_2$ planes and the product over
$j$ is for all other $N=2$ complex planes. For all non ${\bf T}_2\oplus {\bf
T}_4$ ${\bf Z}_N$
orbifolds except ${\bf Z}_6-II-a, b,c$, there is only one $N=2$ complex
plane and so only one complex plane contributing to the
threshold corrections. Thus, for all except the ${\bf Z}_6-II$ cases, either
$${M^2_{string}\over M_X^2}=\alpha_3^{{({{b'}_a}^3-{{b'}_b}^3)}\over
{({{b}_a}-{{b}_b})}}\eqn\twentyon$$
where the $N=2$ complex plane, taken to be the third complex plane, is
a plane where the point group acts as $Z_M$ for $M\not=2$ or
$${M^2_{string}\over M_X^2}=({\alpha_3\tilde\alpha_3})^
{{({{b'}_a}^3-{{b'}_b}^3)}\over
{({{b}_a}-{{b}_b})}}\eqn\twentytw$$
if the $N=2$ complex plane is a plane where the point group acts as $Z_2$.

For all 3 gauge coupling constants to converge to a single value at the same
scale $M_X$ it is necessary that
$${({b'}_3^3-{b'}_2^3)\over({b'}_3^3-{b'}_1^3)}=
{(b_3-b_2)\over (b_3-b_1)}={5\over12}
\eqn\twentythre$$
For this energy scale to be less than $M_{string}$ the sign needed for the
exponent in \twentyon\ or \twentytw\ depends on whether $\alpha_3$ or
$\alpha_3\tilde\alpha_3$ is greater than or less than 1. A numerical study
of the variation of the functions $\alpha_i$ and $\tilde\alpha_i$ of
\te\ and \eleve\ with ${T}_i$ and $U_i$ for the various non ${\bf T}_2\oplus
{\bf T}_4$ orbifolds listed in [\twentytwo]  shows that, although it is
possible for certain
of these functions to attain values greater than 1, they are never much greater
than 1 (never greater than 1.42) On the other hand, the functions
$\alpha_i$ and $\tilde\alpha_i$ can take values very much smaller than 1 for
sufficiently large values of ${T}_i$ and $U_i$. Thus, for
the values of the exponent
 ${{({{b'}_a}^3-{{b'}_b}^3)}\over{({{b}_a}-{{b}_b})}}$ obtained in
practice (numerically less than one) there is no possibility of obtaining a
value of
${M^2_{string}\over M_X^2}$ of the required magnitude except
for the case where $\alpha_3$ or $\alpha_3\tilde\alpha_3$ is less than 1
and raised to
a negative power. We must therefore require
$${{({{b'}_3}^3-{{b'}_2}^3)}\over{({{b}_3}-{{b}_2})}}< 0.\eqn\twentyfou$$
In addition, cancellation  of modular anomalies [\eleven] for the $N=1$
complex planes requires
$${{b'}_3}^i={{b'}_2}^i={{b'}_1}^i, \  i=1,2.\eqn\twentyfiv$$
No further conditions arise from modular anomalies associated with the
$U$ moduli because the modular weights for the $T$ and $U$ moduli
associated with a complex plane are the same state by state.

The conditions to be satisfied for a solution where all 3 gauge coupling
constants converge to a common value at a value of $M_X$ less than
$M_{string}$ for the non ${\bf T}_2\oplus {\bf T}_4$ examples of the ${\bf
Z}_4$,
${\bf Z}_8-II$ and ${\bf Z}_{12}-I$ orbifolds are now identical to the
conditions considered in [\sixteen] for the ${\bf T}_2\oplus {\bf T}_4$
versions of these
orbifolds. The only difference is the values of $T_3$ (and $U_3$ which,
for simplicity, was not included in [\sixteen]) to obtain unification at
$2\times 10^{16}$ GeV would differ because the functions $\alpha_3$ and
$\alpha_3\tilde\alpha_3$ differ from
$({T}_3+\bar {T}_3)ln\vert\eta({T}_3)\vert^4.$
Since no solutions were found to exist in the ${\bf T}_2\oplus {\bf T}_4$ case,
there are still no solutions for these orbifolds.

This leaves only ${\bf Z}_6-II-a, b, c$ as candidate non ${\bf T}_2\oplus {\bf
T}_4$ ${\bf Z}_N$
orbifolds for a successful unification of gauge coupling constants.
For these cases, \twent\  simplifies to
$${M^2_{string}\over
M_X^2}=\alpha_1^{{({{b'}_a}^1-{{b'}_b}^1)}\over{({{b}_a}-{{b}_b})}}
({{\alpha}_3 {\tilde\alpha}_3})^
{{({{b'}_a}^3-{{b'}_b}^3)}\over
{({{b}_a}-{{b}_b})}}\eqn\twentysi$$
where
$$\eqalign{\alpha_1=&({T}_1+\bar { T}_1)\vert\eta({{ T}_1\over2})\vert^4,
\qquad\qquad\alpha_3=({ T}_3+\bar {T}_3)\vert\eta({ T}_3)\vert^2\vert\eta({{
T}_3\over3})\vert^2,\cr
\tilde\alpha_3=&(U_3+\bar U_3)
\vert\eta(U_3)\vert^2\vert\eta(3U_3)\vert^2,\cr &
 \hbox{for} \  {\bf Z}_6-II-a}$$
$$\eqalign{\alpha_1=&({ T}_1+\bar { T}_1)\vert\eta({{ T}_1})\vert^4,
\qquad\qquad\alpha_3=({T}_3+\bar { T}_3)\vert\eta({T}_3)\vert^2\vert\eta({{
T}_3\over3})\vert^2,\cr
\tilde\alpha_3=&(U_3+\bar U_3)
\vert\eta(U_3+2)\vert^2\vert\eta({U_3+2\over3})\vert^2,\cr &
 \hbox{for} \  {\bf Z}_6-II-b}$$
$$\eqalign{\alpha_1=&({T}_1+\bar { T}_1)\vert\eta({{T}_1})\vert^4,
\qquad\qquad\alpha_3=({T}_3+\bar { T}_3)\vert\eta({ T}_3)\vert^2\vert\eta({{
T}_3\over3})\vert^2,\cr
\tilde\alpha_3=&(U_3+\bar U_3)
\vert\eta(U_3)\vert^2\vert\eta(3U_3)\vert^2,\cr &
 \hbox{for} \  {\bf Z}_6-II-c}\eqn\twentyseve$$

Solutions with the 3 gauge coupling constants converging to a single value at
the same scale $M_X$
have been found [\sixteen] for the ${\bf T}_2\oplus {\bf T}_4$ version of the
${\bf Z}_6-II$ orbifold only for the case in which the threshold corrections
are dominated by ${ T}_1$ and do not depend significantly on $T_3$ and $U_3$,
and, as we have argued above, the conditions to be satisfied for a solution to
exist are identical here. Then, it is $\alpha_1$ that determines the value of
$M_{string}^2\over M_X^2$ and the value $ T_1$ required for unification at
$2\times 10^{16}$ GeV is either the same as in the ${\bf T}_2\oplus {\bf T}_4$
case or somewhat larger. (Had solutions existed with $ T_3$ and $U_3$
dominating the threshold corrections, the presence of the factor
$\vert\eta(3U_3)\vert^2$ in $\tilde\alpha_3$ for the
${\bf Z}_6-II-a$ and ${\bf Z}_6-II-c$ cases would have allowed unification with
smaller values of the moduli.)

Convergence of the gauge coupling constants to a common value at $M_X$ may
perhaps be achieved with smaller values of the moduli when the modular
symmetries are broken instead [\nineteen, \twenty] by the presence of discrete
Wilson lines. For example, the choice of Wilson lines given in eqn (60) of the
first reference of [\nineteen] when applied to the ${\bf Z}_3$ plane of the
${\bf Z}_6-II$ orbifold (in the ${\bf T}_2\oplus {\bf T}_4$ version) gives
modular symmetry group $\Gamma_0(3)$ for $ T_1$. It is not known at this time
how to calculate the explicit threshold corrections with discrete Wilson lines.
However, if we conjecture a simple form consistent with the modular symmetries
by employing Dedekind eta functions as in \fou, then we might replace
$\alpha_1$ in \twentyseve\ by
$$\alpha_1=(T_1+\bar T_1)\vert\eta({3{ T}_1})\vert^4.\eqn\twentyeight$$
The orbifold  solution with ${T}_1$ dominating the threshold corrections will
then give convergence of the gauge coupling constants to a common value at
$M_X\approx 2\times 10^{16}$ GeV
with
$$Re {T}_1\approx 8.3.\eqn\twentynin$$
if we make the choice of modular weights displayed in [\sixteen] for which
$${{({{b'}_2}^1-{{b'}_3}^1)}\over{({{b}_2}-{{b}_3})}}=-{1\over4}\eqn\thirty $$
This is to be compared with $Re {T}_1\approx 26$ when the modular symmetry is
unbroken.

Another possible mechanism for convergence of the gauge coupling
constants to a common value to occur at $2\times 10^{16}$ GeV with moderate
values of the moduli is to have the string unification of $SU(3)\times
SU(2)\times U(1)$ gauge coupling constants accompanied by unification to a
gauge group larger than $SU(3)\times SU(2)\times U(1).$
In an earlier paper [\twelve], it has been shown that such a unification of
coupling constants can occur for a number of ${\bf Z}_M\times {\bf Z}_N$
orbifolds (though not for ${\bf Z}_N$ orbifolds) with unified gauge group
$[SU(3)]^3$ or $SO(4)\times SO(6).$

For the case of unification to
$[SU(3)]^3$ with the minimal massless matter content [\twentyfour]
of three copies of $({\bf 3},{\bf 3}, {\bf 1})+({\bf{\bar 3}},{\bf 1},
{\bf{\bar 3}})+({\bf 1}, {\bf{\bar 3}}, {\bf 3})$ to provide the generations
and electroweak Higgses and 2 copies of $( {\bf 1}, {\bf {\bar 3}}, {\bf
3})+({\bf 1}, {\bf 3}, {\bf{\bar 3}})$ providing the $[SU(3)]^3$ breaking
Higgses $H$ and $\bar H$ above the unification scale, and the massless matter
content of the supersymmetric standard model below the unification scale, the
difference of the coefficients ${{b'}_2}^2$ and ${{b'}_3}^i$ for the
$SU(3)\times SU(2)\times U(1)$ threshold corrections is given by
$${{b'}_2}^i-{{b'}_3}^i=6+3\sum _{g=1}^3\Big(n^i_g( {\bf 1}, {\bf{\bar 3}},
{\bf 3})-
n^i_g( {\bf {\bar 3}}, {\bf 1}, {\bf{\bar 3}})\Big)+
3\sum_{f=1}^2\Big(n^i_f(H)+n^i_f(\bar H)\Big).\eqn\thirtyon$$

For the $[SU(3)]^3$ case, all possible choices of modular weights to satisfy
the conditions for the $SU(3)\times SU(2)\times U(1)$ gauge coupling constants
to converge to a common value at a scale less than $M_{string}$ with a single
${ T}_i$ modulus dominating the threshold corrections can be generated using
eqns (27) and (28) of ref. [\twelve] together with a knowledge of all allowed
modular weights of massless states in the twisted sectors of ${\bf Z}_M\times
{\bf Z}_N$ orbifolds [\sixteen, \twentytwo] when the contribution to the
modular weight of the state from the $E_8\times E_8$ algebra $h_{KM}$ satisfies
$h_{KM}\ge {2\over3}.$ We have tabulated in table 2
all the possible values of the exponent $\rho$ in
$${M^2_{string}\over M_X^2}=\Big(({T}_d+\bar {T}_d)\vert\eta({
T}_d)\vert^4\Big)^\rho\eqn\thirtytw$$
where ${T}_d$ is the dominant modulus and
$$\rho={{({{b'}_2}^d-{{b'}_3}^d)}\over{({{b}_2}-{{b}_3})}}\eqn\thirtythre$$
We have also tabulated the values of $Re { T}_d$ which then produce convergence
of the gauge coupling constants to a common value at $2\times 10^{16}$ GeV. Our
notations for the
${\bf Z}_M\times {\bf Z}_N$ orbifolds are as in table 1. It can be seen that
this can be achieved for values of $Re {T}_d$ as small as $3.8$. (The non ${\bf
T}_2\oplus {\bf T}_4$ case need not be considered here because the only ${\bf
Z}_M\times {\bf Z}_N$ Coxeter orbifold for which there are such lattices
[\twentyfive] is
${\bf Z}_2\times {\bf Z}_2$ and there are then no unification solutions
[\twelve] in either the $[SU(3)]^3$ or the
$SO(4)\times SO(6)$ case.)

For the case of unification to $SO(4)\times SO(6)$ with the minimal massless
matter content [\twentysix] of three copies of
$( {\bf { 2}}, {\bf 1}, {\bf{4}})+( {\bf {1}}, {\bf 2},
{\bf{\bar 4}})$ to provide the generations and one copy each of
$( {\bf { 2}}, {\bf 2}, {\bf{1}})+( {\bf {1}}, {\bf 1},
{\bf{6}})$ and $H+\bar H=( {\bf { 1}}, {\bf 2}, {\bf{4}})+( {\bf {1}}, {\bf 2},
{\bf{\bar 4}})$ above the unification scale and the massless matter content of
the supersymmetric standard model below the unification scale, we have instead
$${{b'}_2}^i-{{b'}_3}^i=2\sum _{g=1}^3\Big(n^i_g( {\bf 2}, {\bf{1}}, {\bf 4})-
n^i_g( {\bf {1}}, {\bf 2}, {\bf{\bar 4}})\Big)+
2\Big(n^i( {\bf 2}, {\bf{2}}, {\bf 1})-(n_H^i+n_{\bar H}^i)-n^i( {\bf 1},
{\bf{1}}, {\bf 6})\Big).\eqn\thirtyfou$$

For the  $SO(4)\times SO(6)$ case, the conditions for the $SU(3)\times
SU(2)\times U(1)$
gauge coupling constants to converge to a common value at a scale less than
$M_{string}$ with a single ${T}_i$ modulus dominating the threshold corrections
are eqns. (23) and (24) of ref. [\twelve], and in this case the allowed modular
weights of the twisted sector massless states for  ${\bf Z}_M\times {\bf Z}_N$
orbifolds are those for which $h_{KM}$ satisfies $h_{KM}\ge {5\over8}$ for
$( {\bf { 2}}, {\bf 1}, {\bf{4}}), ( {\bf {1}}, {\bf 2},
{\bf{\bar 4}})$ and $( {\bf { 1}}, {\bf 2}, {\bf{4}})$ and $h_{KM}\ge
{1\over2}$ for $({\bf { 2}}, {\bf 2}, {\bf{1}}) $ and $( {\bf { 1}}, {\bf 1},
{\bf{6}})$.
We have tabulated in table 3, the range of allowed values of the exponent
$\rho$ of \thirtytw\   together with the values of the ${ T}_d$ for which
convergence of the $SU(3)\times SU(2)\times U(1)$ gauge coupling constants to a
common value at $2\times 10^{16} GeV$ is achieved. It can be seen that this can
be achieved for values of $Re {T}_d$ as small as $3.5$

In conclusion, a study has been made of convergence of gauge coupling constants
to a common value at $2\times 10^{16}$ GeV in the context of non ${\bf
T}_2\oplus {\bf T}_4$ ${\bf Z}_N$ orbifolds where the modular symmetries of
threshold corrections are subgroups of $PSL(2, Z)$. The only non ${\bf
T}_2\oplus {\bf T}_4$ orbifolds for which this unification of gauge coupling
constants occurs are those for which it already occurred for the ${\bf
T}_2\oplus {\bf T}_4$ version of the orbifold. In no case can the unification
at
$2\times 10^{16}$ GeV be achieved with smaller values of the moduli than in the
${\bf T}_2\oplus {\bf T}_4$ case. However, when the $PSL(2, Z)$ modular
symmetries are broken instead by discrete Wilson lines, smaller values of the
moduli may be possible, though there is uncertainty as to the detailed form of
the threshold corrections in this case.
We have also considered convergence of gauge coupling constants to a common
value when string unification is accompanied by unification to a gauge group
larger than $SU(3)\times SU(2)\times U(1)$ and have found values of the
dominant ${T}_i$ modulus of around 3, in Planck scale units. will allow
convergence of  $SU(3)\times SU(2)\times U(1)$  gauge coupling constants to
occur at
$2\times 10^{16}$ GeV accompanied by either $[SU(3)]^3$ or $SO(4)\times SO(6)$
unification.

\vskip 0.5cm
\centerline {\bf{ACKNOWLEDGEMENT}}
This work is supported in part by P. P. A. R. C. and the work of S.
Thomas is supported by the Royal Society of Great Britain.
\vskip2cm
\centerline{\bf {Table Captions}}
\noindent
\noindent
{\bf Table}. 1. ${\bf Z}_M\times {\bf Z}_N$ orbifolds. For the point group
generator $\omega$ we display $(\zeta_1, \zeta_2, \zeta_3)$ such that the
action of $w$ on the complex plane orthogonal basis is
$(e ^{2\pi i \zeta_1}, e ^{2\pi i \zeta_2},e ^{2\pi i \zeta_3} )$ and similarly
for the point group generator $\phi.$
\vskip0.2cm
\noindent
{\bf Table}. 2. Values of the exponent $\rho$ in (32) and $Re {T}_d$ for the
various ${\bf Z}_M\times {\bf Z}_N$ orbifolds for threshold corrections
dominated by a single modulus ${T}_d$ and unification to $[SU(3)]^3$ at
$2\times 10^{16}$ GeV.

\vskip0.2cm
\noindent
{\bf Table}. 3. Values of the exponent $\rho$ in (32) and $Re {T}_d$ for the
various ${\bf Z}_M\times {\bf Z}_N$ orbifolds for threshold corrections
dominated by a single modulus ${T}_d$ and unification to $SO(4)\times SO(6)$ at
$2\times 10^{16}$ GeV.
\vskip0.2cm
\vfill\eject
 \vskip 1cm
\vskip 0.5cm
\begintable
Orbifold | Point group generator $\omega \, \, $  |Point group generator $\phi
\, \,$\cr
$Z_2\times Z_2$ | $(1,1,0)/2$| $(0,1,1)/2$ \cr
$Z_4\times Z_2$ | $(1,-1,0)/4$|$(0,1,1)/2$  \cr
$Z_6\times Z_2$ | $(1,-1,0)/6$|$(0,1,1)/2$ \cr
$Z'_6\times Z_2$ | $(1,1,4)/6$|$(0,1,1)/2$ \cr
$Z_3\times Z_3$ | $(1,2,0)/3$|$(0,1,2)/3$
\cr$Z_6\times Z_3$ | $(1,5,0)/6$|$(0,1,2)/3$ \cr
$Z_4\times Z_4$ | $(1,-1,0)/4$|$(0,1,-1)/4$\cr
$Z_6\times Z_6$ |$(1,5,0)/6$|$(0,1,5)/6$
\endtable
\centerline {TABLE 1}
\vskip 2cm
\begintable
Orbifold |Dominant modulus ${T}_d \, \, $|$\rho$|$Re {T}_d$\cr
$Z_3\times Z_3$ |${T}_1$ or ${T}_2$ or ${T}_3$ | $-1$ |$8.2$\cr
$Z_6\times Z_3$ |  ${T}_3$|$-1$  |$8.2$\cr
$Z_4\times Z_2$ | ${T}_1$ or ${T}_2$ |$-0.75$| $10.2$\cr
$Z_4\times Z_4$ | ${T}_1$ or ${T}_2$ or ${T}_3$|$-0.75$ |$10.2$\cr
$Z_6\times Z_2$ | ${T}_1$ or ${T}_2$ |$-0.5$, $-1$, $-1.5$, |$14.3$, $8.2$,
$6.1$,\nr
$Z_6\times Z_3$ |  |$-2$,$-2.5$ or $-3$| $5$, $4.3$ or $3.8$\cr
$Z_6\times Z_6$ | ${T}_1$ or ${T}_2$ or ${T}_3$|$-0.5$,$ -1$, $-1.5$,|$14.3$,
$8.2$, $6.1$,\nr
    |      |  $-2$,$-2.5$ or $-3$      |$5$, $4.3$ or $3.8$\cr $Z'_6\times Z_2$
|${T}_1$ or ${T}_2$ or ${T}_3$|$-0.5$, $-1$|$14.3$ or $8.2$
\endtable
\centerline {TABLE 2}
\vfill\eject
\vskip 0.5cm
\begintable
Orbifold |Dominant modulus ${T}_d \, \, $|$\rho$|$Re {T}_{d} $\cr
$Z'_6\times Z_2$ |${T}_1$ or ${T}_2$ or ${T}_3$ | $18$ values in the range
$-{1\over6}$ to $-{5\over 3}$ |$38$ to $5.7$\cr
$Z_6\times Z_2$ |  ${T}_1$ or ${T}_2$|$38$ values in the range $-{1\over6}$ to
$-{41\over12}$ |$38$ to $3.5$\nr
$Z_6\times Z_3$ |  |and two other values very close to $0$   | \cr
$Z_6\times Z_6$ | ${T}_1$ or ${T}_2$ or ${T}_3$|$38$ values in the range
$-{1\over6}$ to $-{41\over12}$ |$38$ to $3.5$\nr
 |  | and two other values very close to $0$|
\endtable
\centerline {TABLE 3}
\vfill\eject
\refout
\end